# Predicting the Likely Behaviors of Continuous Nonlinear Systems in Equilibrium

Alexander Yeh[1]

MIT Lab. for Computer Science, 545 Technology Sq., NE43-370, Cambridge, MA 02139
phone: (617) 253-3447, electronic mail: ay@xx.lcs.mit.edu

July 21, 1988

## Abstract

This paper introduces a method for predicting the likely behaviors of continuous nonlinear systems in equilibrium in which the input values can vary. The method uses a parameterized equation model and a lower bound on the input joint density to bound the likelihood that some behavior will occur, such as a state variable being inside a given numeric range. Using a bound on the density instead of the density itself is desirable because often the input density's parameters and shape are not exactly known. Current methods tend to either (1) find all the possible system behaviors, but do not tell how likely they are, or (2) just find the *approximate* distribution of possible outcomes, have not been tested on many types of input density bounds, and may have other shortcomings. The new method is called SAB after its basic operations: *split* the input value space into smaller regions, *and* then *bound* those regions' possible behaviors and the probability of being in them. SAB finds rough bounds at first, and then refines them as more time is given.

## 1 Introduction

This paper introduces a method called SAB to predict the likely behaviors of a continuous nonlinear system in equilibrium in which the input values can vary. SAB uses a parameterized equation model and a lower bound on the input joint density to bound the likelihood that one or more state variables stay inside or outside of a given set of numeric ranges.

Current prediction-making methods fit into one of two categories. The ones in the first category find all the possible system behaviors (sometimes including impossible ones), but do not tell how likely the behaviors are. Examples of such methods include various systems which can perform either qualitative reasoning [1], or provide numeric bounds [7]. The ones in the second category approximate the distribution of possible outcomes and may have one or more of the following additional shortcomings: limited forms of acceptable input bounds, runs slowly, will not improve the accuracy of the estimates when given more computation time. Examples of these methods include using moments and Monte-Carlo simulation [2, 4]. A more detailed discussion is at the end of this paper.

A simple example of using SAB involves finding a patient's pulmonary vascular resistance ($PVR$) given the constraint

$$PVR = (PAP - LAP)/CO. \qquad (1)$$

and information on the patient's pulmonary arterial pressure ($PAP$), left atrial pressure ($LAP$) and cardiac output ($CO$). $PVR$ is of interest because a high value indicates that the heart's right ventricle has to work very hard to keep the blood moving through the lungs [5, p. 234]. One threshold condition is $PVR \leq 1.62 mmHg/(l/min)$. Critically ill surgical patients with values above this are less likely to survive [8, p.54-59]. $PAP$, $LAP$, and $CO$ have patient and time dependent values, and are not easy to measure accurately. Table 1 gives some statistics for the

---

[1]This research was supported in part by the National Heart, Lung, and Blood Institute through grant R01-HL33041. The idea for SAB arose after a thought-provoking discussion with Mike Wellman.

|        |       |         | Correlation Coef. |       |        |
| NAME   | MEAN  | STD DEV | PAP   | LAP   | CO     |
|--------|-------|---------|-------|-------|--------|
| PAP    | 23.94 | 3.38    | 1.0   | .861  | .096   |
| LAP    | 15.29 | 3.08    | .861  | 1.0   | -0.044 |
| CO     | 6.49  | 1.20    | .096  | -0.044| 1.0    |

Table 1: $PVR$ Example

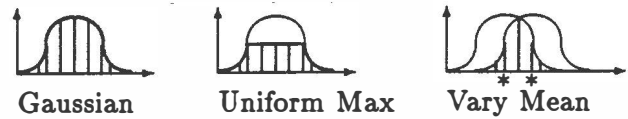

Gaussian     Uniform Max     Vary Mean

Figure 1: 3 Lower Density Bounds

patient of interest, a heart attack victim. The question is, given information on $PAP$, $LAP$, and $CO$ for the patient involved, is $PVR$ at all likely to be above the threshold? If so, one ought to monitor $PVR$.

The numbers are close enough so that the answer is not obvious from looking at Table 1: For example, substituting the mean values into Equation 1 results in $PVR < 1.62$, but increasing $PAP$'s value in the substitution by 3.38 (one standard deviation) while maintaining $LAP$ and $CO$'s values would result in $PVR > 1.62$. However, the latter is not that likely to happen because $LAP$ tends to increase when $PAP$ does (high positive correlation).

So, one has to look at the joint density of $PAP$, $LAP$, and $CO$. Here, one runs into the difficulty that being based on statistics, both the parameter values and the proposed density shape/type are to some extent inaccurate. This difficulty also holds if one uses intuition or heuristics to derive the parameter values and density type.

To get around this difficulty, a user can hypothesize plausible lower bounds on the joint density and let SAB bound the probabilities of satisfying or failing the criteria given each density bound. This can cover the possible variations more completely than using a sample of several possible densities. Ideally, the set of lower density bounds used will cover all the possible variations. In this example, three lower density bounds are considered. One dimensional views of these are in Figure 1, where the areas under the density bounds are marked by vertical lines. The details on getting the results are given later on.

The first "bound" is a regular joint Gaussian density with the parameters listed in Table 1 and is shown in the left diagram of Figure 1. A 1000-sample Monte-Carlo simulation with this bound (a normal probability density) indicates that $PVR > 1.62$ about 20% of the time. SAB analytically bounds this to be between 4% and 57%. This is consistent with the Monte-Carlo simulation and with patient data, where 4 of 17 (23.5%) data points had $PVR > 1.62$.[2]

The second density bound is a

1. joint Gaussian density with the parameters listed in Table 1

2. in which the maximum value is limited to that of a jointly uniform density with the same means and standard deviations.

In other words, the density bound looks like a Gaussian far from the variables' means, but has the low flat top of a uniform density near the means. It is shown in the middle diagram of Figure 1.

---

[2] Here, the data could have been used by itself to answer the question of whether $PVR > 1.62$ is at all likely. SAB is meant to be used when such data is not available.

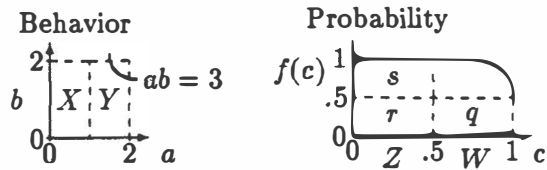

Figure 2: Examples of Splitting

Integrating the bound indicates that it includes $\sim 70\%$ of the probability mass. Using this bound SAB analytically bounds $\Pr(PVR > 1.62)$ to be between 4% and 79%. This is again consistent with the patient data.

The third density bound is the lower bound of a Gaussian density where $CO$'s mean is allowed to be anywhere between 6.20 to 6.78.[3] The right diagram of Figure 1 shows this bound: $CO$'s mean can lie anywhere between the two *'s. The lower density bound is the intersection of the areas under the two Gaussian density curves[4] shown. Integrating the bound indicates that it includes $\sim 65\%$ of the probability mass. Using this bound SAB analytically bounds $\Pr(PVR > 1.62)$ to be between 1% and 76%. This is also consistent with the patient data.

In all three input bound cases, $\Pr(PVR > 1.62) > 1\%$, so $PVR$ should be monitered. Note that the results for each of the input bounds can be tightened. See the end of Section 3 for details.

## 2 SAB: Overview

SAB tightens the probability bound of achieving or failing some criteria by repeatedly selecting a region of possible input values, *splitting* that region into smaller regions $\alpha_i$'s, *and then bounding* both the possible behaviors within the $\alpha_i$'s and the probability of being in the $\alpha_i$'s (using the input probability density bound). SAB marks the $\alpha_i$'s whose possibilities always satisfy or fail the criteria.

Figure 2 shows two examples of splitting. In the one marked *Behavior*, the criterion is $ab < 3$, and the original region is $a, b \in [0, 2]$. In this region $ab \in [0, 4]$, so it sometimes passes and sometimes fails the criterion. Split the region along $a = 1$ into the two sub-regions $X$ and $Y$. In $X$, $a \in [0, 1]$, so $ab \in [0, 2]$, and the criterion is always satisfied. Mark $X$. $a \in [1, 2] \to ab \in [0, 4]$ in $Y$, so $Y$ is not marked. In the example marked *Probability*, $\alpha$, the original region, is $c \in [0, 1]$, and $f(c)$ is a lower bound on probability density at $c$. SAB finds a lower bound on $\Pr(\alpha)$, the probability of being in $\alpha$, of 0.5 (sum areas $q$ and $r$) by multiplying 1, $\alpha$'s length, by 0.5, the lowest value of $f(c)$ in $\alpha$.[5] Split the region at $c = 0.5$ into the two sub-regions $Z$ and $W$. By a method similar to the one above, SAB finds a lower bound on $\Pr(Z)$ of 0.5 (sum areas $r$ and $s$), and a lower bound on $\Pr(W)$ of 0.25 (area $q$). Sum the lower bounds of $\Pr(Z)$ and $\Pr(W)$ to get a new lower bound of 0.75 on $\Pr(\alpha)$.

As hinted by these two examples, *as long as* the bounding method used tends to reduce the range of possibilities as a region of input value space gets smaller, this continued splitting will mark more and more of the value space. And *as long as* the bounding method tends to reduce the gap between a density bound's upper and lower bound in a region as the region gets smaller, the bound on the probability of being in a marked region will improve.

---

[3] The variances, covariances, and other means could also be allowed to vary.
[4] They are the ones with the extreme $CO$ mean values.
[5] Better methods of bounding probabilities are described later.

## 3   $PVR$ Example Revisited

This section re-examines the introduction's $PVR$ example when using the Gaussian density as a "bound" (first density bound). To bound $\Pr(PVR > 1.62)$, SAB looked at the space of inputs (given to SAB as one region):

$$PAP \in [1.0, 88.0], LAP \in [1.0, 88.0], CO \in [1.0, 100].$$

A lower bound on $PVR$, written $lb(PVR)$, is

$$\max(0, [lb(PAP) - ub(LAP)]/ub(CO)) = 0,$$

and an upper bound ($ub(PVR)$) is

$$[ub(PAP) - lb(LAP)]/lb(CO) = 87.0.$$

$PVR$ can be either greater or less than 1.62, so SAB split the space in two along the $CO$ dimension:

$$subspace1: \quad PAP \in [1.0, 88.0], LAP \in [1.0, 88.0], CO \in [1.0, 50.5]$$
$$subspace2: \quad PAP \in [1.0, 88.0], LAP \in [1.0, 88.0], CO \in [50.5, 100.0]$$

SAB then checked and split as appropriate. Regions like

$$PAP \in [20.75, 25.47], LAP \in [15.95, 17.32], CO \in [6.41, 7.19], (PVR \in [0.756, 1.484])$$

where $PVR$ is either always $>$, or $\leq 1.62$, were marked. SAB found lower bounds on the probabilities of being in these marked regions (the one above has a probability $\geq 0.002$).

As SAB recursively splits and checks regions, it tightens the probability bound for satisfying the criteria. When the bound is tight enough, or SAB runs out of time or another resource, it can be stopped. In this example, when SAB was stopped, it gave a lower bound of 0.438 on the probability of being in a passing region (one where $PVR \leq 1.62$), and 0.042 for a failing region ($PVR > 1.62$). If a tighter bound was desired, one could have restarted SAB with the then current set of regions. Since this input joint density bound includes all of the probability mass, SAB can, barring round-off error in the floating point math, get the bound to be arbitrarily tight if given enough computing time. In general, if an input joint density bound includes $n \times 100\%$ of the probability mass, SAB can, barring round-off error, get the bound to have a gap of $1.0 - n$ between the upper and lower figure. So if a density bound includes 70% of the probability mass, the tightest bound SAB could give on the chances of passing some criteria would have a gap of 0.3 between the lower and upper figures (such as a lower bound of 0.6 and an upper bound of 0.9).

## 4   SAB: Details

### 4.1   Main Loop

Perform the following cycle until told to stop:

1. Select the region $\alpha$ with the highest *rank* (see below). SAB can start with either one universal region (as in the example), or any number of predefined regions.

2. What type of region is it?

    (a) *Marked* for being known to always satisfy or fail the given criteria. An example is when a region's $PVR$ range is 0.0 to 1.2 and the criterion is $PVR \leq 1.62$. Here, split the region into two, and using the given density bound, estimate and bound the greatest lower probability bound of being in each of the two sub-regions. Mark them for the same reason as the original region.

    (b) *Unsure*. The region can still either pass or fail the given criteria. An example is when a region's $PVR$ range is 0.0 to 2.0 and the criterion is $PVR \leq 1.62$.
    
    i. If the possibilities of the region ($PVR$'s range in the $PVR$ example) have not been bounded yet, bound them (in the $PVR$ example, use the given formulas for an upper and lower bound on $PVR$). *If* the region should be marked, do so and bound the probability of being in it.
    
    ii. If the possibilities have been bounded, split the region in two. Bound both sub-regions' possibilities, and estimate the greatest lower probability bound of being in each sub-region. *If* a sub-region should be marked, do so and bound that sub-region's probability.

The probability estimations made are just used to suggest the next best step for SAB by helping to rank the sub-regions. They are *not* used as part of any probability bound.

The only overlap allowed between regions is shared borders. No overlap is permitted if the input probability density bound has impulse(s).

## 4.2 Ranking Regions & Estimating Region Probabilities

A region's *rank* estimates how much splitting it will increase the known lower bound on the probability of either satisfying or failing the criteria. An "unsure" (unmarked) region's rank is the estimated greatest lower probability bound (using the given density bound) of being in that region. Estimate as follows:

1. Observe how many input sample points (out of a thousand picked using a "density" which resembles the given input joint density bound) fall within the region. If $> 10$ samples (1%) fall inside, the fraction falling inside is the estimate.

2. If $\leq 10$ samples fall inside, estimate with a formula that quickly, but approximately integrates the density bound in the region. The $PVR$ example uses formula $C_n$ : 3-3 in [9, page 230].

These two parts compensate for each other's weaknesses:

1. The first part is bad for low probabilities because any region $\alpha$ will have large gaps between the sample points within it. So many sub-regions of $\alpha$ will have no sample points even though they may have high values for the lower probability density bound.

2. The second part is bad for high probabilities because the regions involved are either large or probably contain a complicatedly shaped part of the density bound.[6] The integration formulas only work well when a region's section of the density bound is easily approximated by a simple polynomial.

---

[6] Most of the common probability densities only have complicated shapes where the density values are high. I am assuming that this complication will be reflected in the corresponding part of the bound.

A marked region's rank is the gap between the estimated greatest lower probability bound of being in the region and the known lower bound on that probability. This works better than the gap between the upper and lower bounds on the greatest lower probability bound because SAB often finds very loose upper bounds, while the estimates are usually accurate.

### 4.3 Bounding Region Probabilities

The basic way SAB finds a lower bound on the probability of being in a region is to multiply the region's volume[7] by its minimum probability density lower bound value (found by the bounding mechanism described below). I derived the $PVR$ example's first density bound expression (a Gaussian density) by taking the density parameters (Table 1) and substituting them into the general form for a Gaussian density. After some simplification, I got (numbers rounded-off):

$$0.01033 \exp(-0.01323(13.70P^2 - 26.09PL - 10.36PC + 16.42L^2 + 10.77LC + 28.08C^2))$$

where $P = (PAP - 23.94)$, $L = (LAP - 15.29)$, and $C = (CO - 6.487)$.

To help tighten this bound, SAB tries to use any monotonicity and/or convexity present in the region's part of the density bound in the following manner:[8]

Let $f(x_1, \ldots, x_n)$ be the probability density and within a region $\alpha$ let $x_i$ range between $l_i$ and $h_i$. The probability of being in $\alpha$ is

$$F = \int_{l_n}^{h_n} \cdots \int_{l_1}^{h_1} f(x_1, \ldots, x_n) dx_1 \ldots dx_n.$$

If $\partial f / \partial x_1$ is always $> 0$ in $\alpha$, then

$$F \geq [\prod_{i=1}^{n}(h_i - l_i)][(\min_* f(l_1, x_2, \ldots, x_n)) + (\min \frac{\partial f}{\partial x_1}(x_1, \ldots, x_n))(\frac{h_1 - l_1}{2})],$$

where the minimization of $f$ is over the $x_2$ through $x_n$ values within $\alpha$ ($\min_*$ means that $x_1$ is NOT part of the minimization) and the minimization of $\frac{\partial f}{\partial x_1}$ is over the $x_1$ through $x_n$ values within $\alpha$. This bound is tighter than the basic lower bound:

$$[\prod_{i=1}^{n}(h_i - l_i)][\min f(x_1, \ldots, x_n)].$$

Similar expressions can be derived for the other variables and for when $\partial f / \partial x_i < 0$.

If $\partial^2 f / \partial x_1^2$ is always $\leq 0$ in $\alpha$ (convex down), then

$$F \geq [\prod_{i=1}^{n}(h_i - l_i)][(\min_* f(l_1, x_2, \ldots, x_n)) + (\min_* f(h_1, x_2, \ldots, x_n))]/2,$$

where the minimizations of $f$ are over the $x_2$ through $x_n$ values within $\alpha$. This bound is also tighter than the basic one. See Figure 3 for the one dimensional case: the ∩ curve is the density, the area under the diagonal line is $F$'s new lower bound, and the area under the horizontal line is the original bound. Similar expressions can be derived for the other variables.

Several methods exist to integrate a region's probability density bound, including Monte-Carlo [3] and quadrature (numeric integration) methods [9]. These cannot truly bound the integration error because they only take numeric samples at particular points.

---

[7]For a region $\alpha$, let its variables $x_i$ ($i = 1 \ldots n$) range between $l_i$ and $h_i$. Then $\alpha$'s volume is $\prod_{i=1}^{n}(h_i - l_i)$. SAB only deals with $n$-dimensional rectangular regions.

[8]Derivations forthcoming.

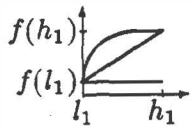

Figure 3: 1-D Convex Density and Lower Bound

### 4.4 Splitting Regions

SAB may split a selected region $\alpha$ in either step 2a or step 2(b)ii. In either, SAB picks a variable in $\alpha$ to split along and then bisects $\alpha$. Select the variable as follows: in step 2(b)ii, find the one with the largest difference between its upper and lower bound within the region, normalized by its standard deviation. In step 2a, find the one with the largest apparent variation in the density's slope with respect to it.

### 4.5 Finding Numeric Bounds

Many of SAB's parts need to bound expressions. For algebraic expressions (the type in the models to be used), perfect bounding algorithms have not been built. The type of algorithm used here will find bounds that indicate the truly unachievable,[9] but may not be the tightest possible. Example: saying that $x < 7$, when in fact, $x < 3$. I have implemented an augmented version of bounds propagation [7]. It does the following interval arithmetic [6]:

- Bound an operation's result using bounds on the operation's operands. Example: $ub(a + b) \leq ub(a) + ub(b)$.

- Bound an operand using bounds on an operation's result and the operation's other operands. Example: $ub(a) \leq ub(a + b) - lb(b)$.

The bounder examines expressions and updates bounds with these operations. It *iterates* over the expressions until every one that might produce a change has been examined at least once and all the recent bound changes are below a certain threshold.

## 5 Discussion

This paper introduces a method called SAB which uses a lower bound on the input joint density to analytically bound the likelihood of some possible behavior.

SAB entails much computation. When possible, first estimate the probability with some approximation method like moments and then use SAB to insure that the probability is within certain bounds. Also, once regions are made in response to one set of criteria, they can be reused when examining other sets. This will cut down much of the computation for the remaining sets.

Monte-Carlo[2, 4] is probably the most similar method to SAB. When using certain densities (such as Gaussian) for which fast variate generation routines exist or limited linear combinations (mixtures of scaled versions) of such densities[10] as an input bound, Monte-Carlo will run faster then SAB. But for other density bounds, such as the second and third density bounds given in the introduction's $PVR$ example, I am not so sure. Using Monte-Carlo for these will require combining numeric bounding techniques and the acceptance-rejection method of variate generation. The former will find the density bound upon which the latter will use as the accept/reject threshold. The latter is slow because it often rejects many generated random numbers before accepting one. Either way, Monte-Carlo gives estimates based on random numbers, not analytical bounds.

---
[9] In practice, the accuracy of this may be limited by round-off error.

[10] For example: two times a certain Gaussian density plus one third some exponential density.

Another method dealing with probabilities is using moments[2]. This method can only deal with an input bound that is a mixture of scaled densities or is a density where one or more of the parameter values are bounded within a certain interval (like the third density bound in the introduction's $PVR$ example). Handling the latter will require numeric bounding techniques to bound the various moments (mean, variance, etc.). The estimates are not analytic bounds and unlike Monte-Carlo, the accuracy of the *estimates* of the various likelihoods do not improve as more computation time is given.

The combining of numeric bounding and either Monte-Carlo or moments described above has not been tried yet and doing so is part of future work. Also to be done is finding various types of input density bounds and determining when which should be used.

Future work on SAB itself includes (1) testing it by using it on a full scale cardiovascular model to examine the effects of heart disease therapies and (2) expanding it to deal with upper input density bounds and to more quickly bound a variable's mean, variance, median, 90% confidence interval, etc. I will also explore splitting a region at the selected variable's median value (or some approximation) within the region. This can handle infinite intervals (bisection cannot), which permits an initial region where each variable is within the all-inclusive range of $[-\infty, \infty]$.

Until now, only approximate likelihoods of possible system behaviors with no known analytic error bounds have been found and despite uncertainty in input density shapes and parameter values, bounds on input densities have not been really utilized. SAB, and the line of work from which it comes, is designed to remedy this.

# References

[1] Daniel G. Bobrow, editor. *Qualitative Reasoning about Physical Systems*. MIT Press, 1985.

[2] Gerald Hahn and Samuel Shapiro. *Statistical Models in Engineering*. John Wiley and Sons, Inc., 1967.

[3] J. M. Hammersley and D. C. Handscomb. *Monte Carlo Methods*. Barnes and Noble, Inc., 1965.

[4] Mark Johnson. *Multivariate Statistical Simulation*. John Wiley and Sons, New York, 1987.

[5] Dean T. Mason. *Congestive Heart Failure, Mechanisms, Evaluation, and Treatment*. Dun-Donnelley, New York, 1976.

[6] H. Ratschek and J. Rokne. *Computer Methods for the Range of Functions*. Halsted Press: a division of John Wiley and Sons, New York, 1984.

[7] Elisha P. Sacks. Hierarchical reasoning about inequalities. In *Proceedings of the National Conference on Artificial Intelligence*, pages 649–654. American Association for Artificial Intelligence, 1987.

[8] William C. Shoemaker. Physiology, monitoring and therapy of critically ill general surgical patients. In William C. Shoemaker and Edward Abraham, editors, *Diagnostic Methods in Critical Care*, chapter 4, pages 47–86. Marcel Dekker, Inc., New York, 1987.

[9] A. H. Stroud. *Approximate Calculation of Multiple Integrals*. Prentice-Hall, Inc., 1971.